\documentclass[10pt, conference, letterpaper]{IEEEtran}
\IEEEoverridecommandlockouts

\usepackage{cite}
\usepackage{amsmath,amssymb,amsfonts}
\usepackage{algorithmic}
\usepackage{graphicx}
\usepackage{textcomp}
\usepackage{xcolor}
\usepackage{multirow}
\def\BibTeX{{\rm B\kern-.05em{\sc i\kern-.025em b}\kern-.08em
    T\kern-.1667em\lower.7ex\hbox{E}\kern-.125emX}}
\begin{document}

\title{Information-Entropy-Driven Fault Propagation Modeling for Probabilistic Network Performance Prediction}



\author{
\IEEEauthorblockN{
1\textsuperscript{st} Lusha Mo,
2\textsuperscript{nd} Fengxiao Tang\textsuperscript{*},
3\textsuperscript{rd} Xiaonan Wang,
4\textsuperscript{th} Ming Zhao
}
\IEEEauthorblockA{
\textit{School of Computer Science and Engineering,} \\
\textit{Central South University,} 
Changsha, China \\
\{molusha, tangfengxiao, wangxiaonan, meanzhao\}@csu.edu.cn \\
\textsuperscript{*} Corresponding Author
}
}

\maketitle

\begin{abstract}
Network faults can trigger cascading effects that cause abrupt and nonstationary performance degradation. Existing learning-based performance predictors mainly focus on normal operation or treat fault-induced topology and routing changes as static inputs, and typically produce deterministic point estimates. They overlook fault-propagation dynamics and uncertainty in performance evolution. The predefined-rule and purely data-driven propagation models lack a unified representation of fault definition, propagation mechanism, and impact quantification. Additionally, generic denoisers in conditional diffusion models fail to incorporate fault propagation into uncertainty modeling. To address these limitations, we propose an information-entropy-driven fault propagation paradigm (IEFP) that characterizes fault propagation via relative entropy, mutual information and transfer entropy. We then design a fault-aware graph message-passing mechanism that propagation contexts modulate network representation learning. We further develop FEMNet, which employs this mechanism as a tailored denoiser within a conditional diffusion model to enable probabilistic network performance prediction under complex fault scenarios. Compared with the strongest baselines, IEFP improves fault-prediction performance, while FEMNet reduces errors in both point and probabilistic KPI prediction. 

\end{abstract}

\begin{IEEEkeywords}
Network fault propagation, probabilistic network performance prediction, information-entropy modeling, conditional diffusion model, graph message passing.
\end{IEEEkeywords}

\section{Introduction}
Network performance prediction estimates how key performance indicators (KPIs) evolve under different network conditions. Faults can trigger link disruptions, route reconfiguration, and queue buildup, causing cascading effects and quality-of-service degradation~\cite{song2015dynamic,varbella2024powergraph}. The deterioration is abrupt, nonstationary, and driven by structural changes in the network. Accurate and rapid assessment of fault impacts on network KPIs is therefore essential for proactive degradation detection, automated control, and intelligent maintenance, as well as for robustness evaluation and reliability enhancement during network design and deployment.

\begin{figure}
\centerline{\includegraphics[width=\columnwidth]{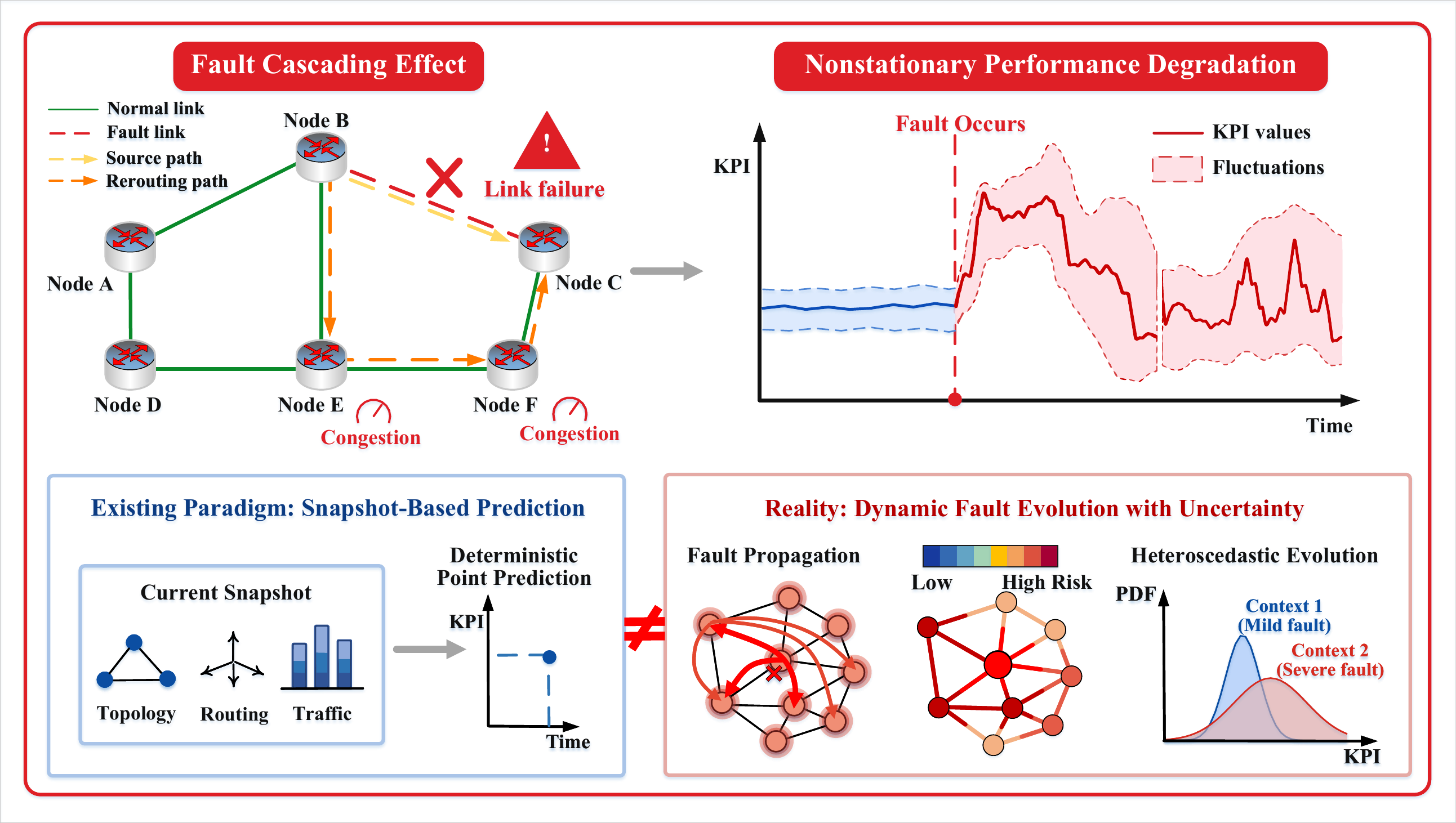}}
\caption{Fault cascading and its impact on network performance. A local link failure can trigger rerouting and cascading congestion (top left), leading to nonstationary KPI degradation (top right). Conventional snapshot-based deterministic prediction (bottom left) cannot represent the fault propagation and context-dependent, heteroscedastic KPI distributions(bottom right).}
\label{fig:motivation}
\end{figure}

Learning-based predictors\cite{yang2022deepqueuenet} are prominent, which capture complex relationships among topology, routing, and traffic to predict node-, link-, and end-to-end performance. However, most assume normal operation under fixed configurations. Even when link failures are considered\cite{ferriol2023routenet}, the resulting topology, route, and other reachability variations are merely encoded as updated static inputs. Such updates omit temporal dependencies and the accumulated effects of preceding faults. Performance prediction thus remains a snapshot-based mapping from network state to deterministic KPIs, as shown in Fig.~\ref{fig:motivation}, without explicitly modeling how faults propagate and how the process shapes KPI evolution. This limitation becomes critical under complex faults scenarios.

A central challenge is to accurately model fault propagation. Existing approaches either rely on predefined rules\cite{lin2023dynamic,zhou2025reliability} which generalize poorly to heterogeneous dependencies and stochastic triggers, or fit propagation patterns from data\cite{tang2024semi} without characterizing the persistent effect of historical faults on future fault occurrence. They lack a unified account of three fundamental questions: what constitutes a fault, how it propagates, and how its impact should be quantified. In fact, a fault implies that a component deviates from normal operation, and the abnormality may propagate along dependencies and alter the uncertainty and evolution of other components' future states. Information theory provides a natural language for characterizing uncertainty, coupling strength, and directional influence\cite{wang2024cumulative}, motivating an information-entropy\cite{zhang2023new,bauer2006finding} view for modeling fault propagation.

Furthermore, performance prediction must also capture how fault propagation affects KPI evolution. Under fault conditions, node and link interactions depend not only on topology, routing, and traffic but also on the accumulated effects of historical faults, which can persistently raise abnormality risks in dependent components. Since fault propagation changes the importance, reliability, and effective range of information exchanged over the graph, its context should dynamically regulate message passing instead of serving as static feature.

Another difficulty arises from the pronounced uncertainty and heteroscedasticity of network performance evolution under fault conditions. Fault propagation may induce complex stochastic variations through multiple joint factors, such that similar network states and fault contexts may yield markedly different degradation patterns and fluctuation magnitudes. Performance prediction should therefore move beyond deterministic point estimates to characterize conditional KPI distributions. Conditional diffusion models\cite{zhang2023adding} offer a natural mechanism for this purpose. However, the generic denoisers do not model fault propagate and its effects on future states\cite{li2025diffgraph}. Probabilistic prediction under fault conditions requires a fault-aware denoising architecture.

We address these challenges by jointly modeling fault propagation and uncertainty in KPI evolution. We first establish IEFP, an information-entropy-driven fault propagation paradigm to estimates fault risks for nodes and links. Specifically, IEFP uses relative entropy, mutual information, and transfer entropy to characterize abnormality levels, inter-node coupling strength, and cross-component influence from historical to future states, respectively. We then design a fault-aware graph message passing mechanism to incorporate propagation information into graph representation learning, where predictive risks gate messages, adjust attention weights, and mask weakly coupled edges. Using the mechanism as denoising network, we develop FEMNet, a probabilistic network performance prediction model under complex fault scenarios, which introduces conditional diffusion to generalize deterministic point prediction to its probabilistic counterpart. Our main contributions are as follows.

1) We propose IEFP, an information-entropy-driven fault propagation paradigm that interprets fault propagation as abnormal information transmission, accumulation and redistribution over networks, providing a new theoretical perspective for fault propagation analysis in complex network scenarios.

2) To effectively incorporate propagation information into performance prediction, we design a fault-aware graph message passing mechanism, in which predictive fault risks modulates node message and update dynamics.

3) We develop FEMNet for probabilistic network performance prediction under complex fault scenarios. By introducing conditional diffusion model, FEMNet transforms performance prediction from deterministic point regression task into conditional distribution generation task.

4) We evaluate IEFP and FEMNet under complex fault scenarios in terms of point-prediction accuracy, fault-anticipation capability, and probabilistic-prediction quality. Results demonstrate consistent improvements over the evaluated baselines.

\section{Related Work}
\subsection{Network Performance Prediction}
Among accurate, lightweight learning-based network models, RouteNet\cite{rusek2020routenet} based on graph neural network (GNN), models links and traffic paths to predict source-destination KPIs. xNet\cite{huang2023xnet} and RouteNet-Erlang\cite{ferriol2022routenet} incorporate queue characteristics and specialized graph message passing. HTNet\cite{zhou2023htnet}, DGAT\cite{yu2023digital}, and EAGLE\cite{liu2023eagle} capture temporal behavior, attention weights, and bandwidth-aware representations. RouteNet-Fermi~\cite{ferriol2023routenet} accepts topology and routing changes after link failures as inputs but omits fault interaction and diffusion. A recent analysis~\cite{bernardez2025ordered} interprets RouteNet as high-order topological modeling that captures path and scheduling order, while RouteNet-Gauss~\cite{guemes2026routenet} adds hardware-testbed data and configurable temporal granularity. These methods rely on configurations or state snapshots and do not model fault propagation and its persistent effects on performance evolution.

\subsection{Fault Propagation Modeling}
Fault propagation has been studied using Bayesian networks\cite{zheng2024algorithms}, fault trees, cascading models (e.g., threshold models, load-capacity models\cite{zhu2025small}, and interdependent network\cite{roth2025cascading}) and diffusion models. State explosion and predefined rules limit their ability to capture probabilistic dynamics, heterogeneous dependencies, and temporal interactions. For example, HIC model\cite{xiang2024predicting} combines stochastic Markovian diffusion with physics-based concepts. Probabilistic graphical models offer interpretability and causal expressiveness(e.g., Causal model\cite{ghosh2024cascading}), but their structure learning and inference remain challenging at scale. Data-driven methods, including machine learning\cite{tang2024semi}, Graph Hawkes processes\cite{cai2022thps,linderman2014discovering}, and GNN-based models (e.g., GPIAN\cite{yang2025lightweight}), fit numerical variations without explicitly representing fault triggering and propagation. Existing methods lack a unified representation of how fault impacts traverse complex networks and persist over time. Information theory concerned with information, uncertainty and communication, provides a natural foundation for such theoretical framework.

\subsection{Conditional Diffusion Model}
Diffusion models\cite{hu2024anomalydiffusion,sun2025anomaly} have expanded from image generation to probabilistic time-series and graph prediction. TimeDiff\cite{shen2023non} applies conditional diffusion to non-autoregressive time-series prediction, and DiffSTG\cite{wen2023diffstg} combines diffusion models with spatiotemporal graph modeling for probabilistic forecasting. LGD\cite{zhou2024unifying} formulates graph regression and classification tasks into a generic conditional generation framework. ReDiSC\cite{li2025redisc} reparameterized masked diffusion for structured node classification. SimDiff\cite{ding2026simdiff} improves point estimation using normalization-independent diffusion and a median-of-means estimator. These studies demonstrate the strong potential of diffusion models for uncertainty modeling in network performance prediction. However, generic denoisers lack fault-propagation mechanisms and thus cannot fully capture fault-aware network dynamics.

\section{Problem Formulation}
\label{sec:preliminaries_problem}
This section reviews the required information-theoretic  quantities and conditional diffusion foundations, and formulates probabilistic network performance prediction under complex fault conditions.

\subsection{Preliminaries}
\textbf{Definition 1 (Information-theoretic quantities).}
Let $P$ and $Q$ be probability distributions on $\Omega\subseteq\mathbb R^d$ with density functions $p(x)$ and $q(x)$, respectively, where $q(x)>0$ whenever $p(x)>0$. Their relative entropy is
\begin{equation}
D_{\mathrm{KL}}(P\|Q)
\equiv
\int_{\Omega}p(x)\log\frac{p(x)}{q(x)}\,dx,
\label{eq:relative_entropy}
\end{equation}
which is $+\infty$ if the support condition fails. For random elements $B$, $C$, and $D$ on standard Borel spaces with joint law $P_{BCD}$, mutual information and conditional mutual information are 
\begin{equation}
\begin{aligned}
I(B;C)
&\equiv D_{\mathrm{KL}}
\!\left(P_{BC}\middle\|P_B\otimes P_C\right),\\
I(B;C\mid D)
&\equiv \mathbb{E}_{P_D}\!\left[
D_{\mathrm{KL}}\!\left(
P_{BC\mid D}\middle\|
P_{B\mid D}\otimes P_{C\mid D}
\right)\right].
\end{aligned}
\label{eq:mutual_information}
\end{equation}
The regular conditional laws in~\eqref{eq:mutual_information} are understood $P_D$-almost surely. Let $B_t^{(k_B)}$ and $C_t^{(k_C)}$ denote the source and target history vectors, respectively. The context-conditioned transfer entropy from $B$ to $C$ is
\begin{equation}
\mathrm{TE}_{B\to C\mid D}(t)
\equiv I\!\left(
C_{t+1};B_t^{(k_B)}
\mid C_t^{(k_C)},D_t
\right),
\label{eq:conditional_transfer_entropy}
\end{equation}
which measures directed predictive dependence under the specified history and context. 

\textbf{Definition 2 (Conditional diffusion model).} A conditional diffusion model is a probabilistic generative model that learns $p(\mathbf{y}_0\mid\mathbf{c})$, where $\mathbf{c}$ denotes conditional information such as labels, descriptions, attributes, or states. The forward diffusion process gradually injects Gaussian noise into data samples:
 \begin{equation}
 \begin{aligned}
q(\mathbf{y}_{1:T}\mid\mathbf{y}_0)
&=\prod_{\tau=1}^T\mathcal{N}\left(\mathbf{y}_\tau;\sqrt{1-\beta_\tau}\mathbf{y}_{\tau-1},\beta_\tau\mathbf{I}\right),
\end{aligned}
 \end{equation}
where $\tau$ is diffusion step and $\beta_\tau$ is a predefined noise-schedule coefficient. Let $\alpha_\tau=1-\beta_\tau$ and $\bar{\alpha}_\tau=\prod_{s=1}^{\tau}\alpha_s$. Then
\begin{equation}
\mathbf{y}_\tau
=\sqrt{\bar{\alpha}_\tau}\mathbf{y}_0
+\sqrt{1-\bar{\alpha}_\tau}\boldsymbol{\epsilon},
\quad \boldsymbol{\epsilon}\sim\mathcal{N}(\mathbf{0},\mathbf{I}).
\label{eq:forward_marginal}
\end{equation}

Starting from pure noise, the reverse denoising process gradually recovers target samples satisfying condition $\mathbf{c}$:
\begin{equation}
p_\theta(\mathbf{y}_{0:T}\mid\mathbf{c})=p(\mathbf{y}_T)\prod_{\tau=1}^Tp_\theta(\mathbf{y}_{\tau-1}\mid\mathbf{y}_\tau,\mathbf{c}),
\end{equation}
where the transition $p_\theta(\mathbf{y}_{\tau-1}\mid\mathbf{y}_\tau,\mathbf{c})$ is parameterized by a neural network and explicitly depends on $\mathbf{c}$.

The model learns to restore the original data by predicting the injected noise. Training objective is to minimize the loss between the actual and predicted noise:
\begin{equation}
\min_\theta\mathcal{L}(\theta)=\mathbb{E}_{\mathbf{y}_0,\epsilon,\tau,\mathbf{c}}\left[\left\|\epsilon-\epsilon_\theta(\mathbf{y}_\tau,\tau,\mathbf{c})\right\|_2^2\right].
 \end{equation}

\subsection{Problem Definition}
Considering a communication network, fault events may dynamically alter network topology, resource availability, routing behavior, and traffic evolution. Their effects may propagate across dependent entities and introduce substantial performance uncertainty. We aim to predict the probability distributions of network KPIs at the next timestamp under complex fault scenarios.

A fault scenario is complex when performance is affected by coupled faults rather than one isolated perturbation, including at least one of: 1) concurrent faults; 2) temporally correlated faults within the observation window whose effects propagate through network dependencies; or 3) faults differing in type, duration, or severity that jointly affect multiple entities.

We represent the network as a heterogeneous graph $G=(V,E,A)$, where node sets $V$ represent queues, links, and flows; edge sets $E$ encode their dependencies; and $A$ is the adjacency matrix. For node $v$, let $\mathbf{z}_v\in\mathbb{R}^{d_z}$ be its static configuration and $X_v(t)\in\mathbb{R}^{d_x}$ its time-varying feature vector, with observation $\mathbf{x}_v(t)$. Examples include a queue's buffer limit in $\mathbf{z}_v$ and current occupancy in $\mathbf{x}_v(t)$, or a link's maximum load in $\mathbf{z}_v$ and current bandwidth in $\mathbf{x}_v(t)$. With $\mathbf{x}_v^{0:t}=(\mathbf{x}_v(0),\ldots,\mathbf{x}_v(t))$, the network attributes are
\begin{equation}
\mathcal{Z}=\left\{\left(\mathbf{z}_v,\mathbf{x}_v^{0:t}\right):v\in V\right\}.
\end{equation}
The traffic states are $\mathcal{S}=\{\mathbf{s}_f(t)\in\mathbb{R}^{d_f}\}_{f\in \mathcal F}$, where $\mathbf{s}_f(t)$ is flow $f$'s current feature vector. Faults within observation window $[0,t]$ form the event sequence
\begin{equation}
\mathcal{E}_{0:t}=\left\{\varepsilon_k=\left(\mathrm{type}_k,\mathrm{loc}_k,t_k^\mathrm{start},t_k^\mathrm{end}\right)\right\}_{k=1}^K,
\end{equation}
whose entries denote the fault type, affected entity, start time and end time of the $k$-th fault event, respectively.

Given condition $\mathcal{C}= (G,\mathcal{Z},\mathcal{S},\mathcal{E}_{0:t})$, we estimate each node's next-step KPI distribution:
\begin{equation}
\mathcal{Y}=\left\{y_v(t+1)\sim p_v(\cdot\mid G,\mathcal{Z},\mathcal{S},\mathcal{E}_{0:t})\right\}_{v\in V},
\end{equation}
where $y_v(t+1)$ and $p_v(\cdot)$ denote node $v$'s target KPIs and their conditional distribution. KPIs include delay, throughput, queue length, packet loss, and related metrics.

Therefore, We learn a mapping
\begin{equation}
f_\Theta:(G,\mathcal{Z},\mathcal{S},\mathcal{E}_{0:t})\to\mathcal{Y},
\end{equation}
that enables the predicted distributions accurately characterize stochastic performance evolution under the joint effects of traffic dynamics, resource contention, and fault propagation.

For training, applying~\eqref{eq:forward_marginal} to the normalized KPI target $\mathbf{y}_0$ gives $\mathbf{y}_\tau$. We optimize
\begin{equation}
\min_\theta\mathcal{L}_{\mathrm{diff}}(\theta)
=\mathbb{E}_{\mathbf{y}_0,\boldsymbol{\epsilon},\tau,\mathcal{C}}
\!\left[\left\|\boldsymbol{\epsilon}
-\boldsymbol{\epsilon}_\theta(\mathbf{y}_\tau,\tau,\mathcal{C})
\right\|_2^2\right].
\label{eq:kpi_diffusion_loss}
\end{equation}
where $\epsilon_\theta$ is the denoising network. By minimizing $\mathcal{L}_{\mathrm{diff}}$, the model approximates the KPI distribution conditioned on network state and fault information.

\section{Information-Entropy-Driven Fault Propagation Paradigm}
\label{sec:propagation_paradigm}
A fault indicates a component's departure from normal operation and injects abnormal information into the system. This information propagates through topological dependencies, resource interactions, and protocol coupling, altering other components' uncertainty and dynamics. IEFP model fault propagation via three information-theoretic quantities. Relative entropy measures deviation of observation distribution from normal operation and defines fault severity. Mutual information decomposes fault information into configuration vulnerability, local historical degradation, and propagated influence. Transfer entropy quantifies the directional incremental information that one component's history provides about another's fault state.

We define
\begin{equation}
\mathcal{H}_v(t)=\bigl(X_v(t-L+1),\ldots,X_v(t-1),X_v(t)\bigr)
\end{equation}
as node $v$'s length-$L$ history up to time $t$, and let $p_v^{(N)}(X)$ be the reference distribution estimated from normal samples.

\textbf{Theorem 1 (Asymptotic Equipartition Property, AEP).} If the random process $\{X_v(t)\}$ is stationary and ergodic, then as $n\to\infty$,
\begin{equation}
-\frac{1}{n}\log P\left(X_v(1),\ldots,X_v(n)\right)\to \bar{H}(X_v).
\end{equation}
Thus, average information of a long normal sequence concentrates around entropy rate $\bar{H}(X_v)$, and normal observations lie in the typical set with high probability. A state unlikely under the normal distribution carries information unexplained by the nominal operating regime.

Let $\hat{p}_v^t(X)$ be the observation distribution constructed from node $v$'s current feature. We define its anomaly information quantity as the relative entropy between the current observation and normal reference distributions:
\begin{equation}
S_v(t)=D_{\mathrm{KL}}\!\left(\hat{p}_v^t(X)\,\|\,p_v^{(N)}(X)\right).
\end{equation}
Larger $S_v(t)$ indicates greater distributional departure from normal generation mechanism and a more severe anomaly. A fault occurs when this information exceeds the system tolerance, yielding the binary state
\begin{equation}
F_v(t)=
\begin{cases}
1, & S_v(t)\geq\lambda_v,\\
0, & S_v(t)<\lambda_v,
\end{cases}
\end{equation}
where $\lambda_v>0$ is node $v$'s learned tolerance threshold. The entropy-marked fault-event history is
\begin{equation}
\mathcal{A}_v(t)=\{(t_k,S_v(t_k)):t_k\leq t,\,F_v(t_k)=1\},
\end{equation}
which records each fault’s timestamp and anomaly magnitude.

\textbf{Theorem 2 (Conditional reduction of entropy)}. For random variables $B$, $C$, and $D$, we have:
\begin{equation}
H(C\mid B,D)\leq H(C\mid B),
\end{equation}
with equality iff $C \perp\!\!\!\perp D \mid B$, i.e., $D$ provides no additional information about $C$ beyond what $B$ already contains.

Faults propagate because nodes are coupled by shared information, physical or logical constraints, routing, and resource competition. A fault at node $u$ changes its output distribution:
\begin{equation}
P(X_u(t)\mid F_u(t)=1)\neq P(X_u(t)\mid F_u(t)=0).
\end{equation}
This output enters adjacent nodes' state dynamics and may change their posterior fault risk: 

\begin{equation}
\begin{aligned}
&P\!\left(F_v(t+1)=1\mid\mathcal{H}_v(t),\mathcal{H}_u(t),\mathbf{z}_v\right)
\\&\quad \neq
P\!\left(F_v(t+1)=1\mid\mathcal{H}_v(t),\mathbf{z}_v\right).
\end{aligned}
\end{equation}

Information-theoretically, faults in complex systems are conditionally stochastic. Anomalous information from neighbors’ history propagates along dependency edges, adding condition-dependent information about the target’s fault state. By Theorem~2:
\begin{equation}
\begin{aligned}
&H\!\left(F_v(t+1)\mid\mathcal{H}_v(t),\mathbf{z}_v\right)\\
&\quad >
H\!\left(F_v(t+1)\mid\mathcal{H}_v(t),\mathcal{H}_u(t),\mathbf{z}_v\right).
\end{aligned}
\end{equation}

We quantify this directed historical incremental information by transfer entropy:
\begin{equation}
\begin{aligned}
\mathrm{TE}_{u\rightarrow v}
&=
I\!\left(
F_v(t+1);
\mathcal{H}_u(t)
\,\middle|\,
\mathcal{H}_v(t),\mathbf{z}_v
\right)                                                   \\
&=
H\!\left(
F_v(t+1)
\,\middle|\,
\mathcal{H}_v(t),\mathbf{z}_v
\right)                                                   \\
&\quad -
H\!\left(
F_v(t+1)
\,\middle|\,
\mathcal{H}_v(t),\mathcal{H}_u(t),\mathbf{z}_v
\right).
\end{aligned}
\label{eq:transfer_entropy}
\end{equation}
A larger $\mathrm{TE}_{u\to v}$ indicates a stronger statistical association between neighbor $u$'s history and node $v$'s fault state, and hence stronger propagation. 

\textbf{Theorem 3 (Chain rule of mutual information).} For random variables $B$, $C$, and $D$,
\begin{equation}
I(D;B,C)=I(D;B)+I(D;C\mid B).
\end{equation}
Repeatedly applying Theorem~3 decomposes the total statistical association between node $v$'s fault state and the available configuration, local history, and neighborhood history:

\begin{equation}
\begin{aligned}
&I\!\left(
F_v(t+1);
\mathcal{H}_v(t),
\mathcal{H}_{\mathcal{N}(v)}(t),
\mathbf{z}_v
\right)                                                   \\
&\quad =
\underbrace{
I\!\left(F_v(t+1);\mathbf{z}_v\right)
}_{I_v^{(\mathrm{conf})}}                                  \\
&\qquad +
\underbrace{
I\!\left(
F_v(t+1);
\mathcal{H}_v(t)
\,\middle|\,
\mathbf{z}_v
\right)
}_{\substack{
I_v^{(\mathrm{local})}: \text{local term}
}}                                                         \\
&\qquad +
\underbrace{
I\!\left(
F_v(t+1);
\mathcal{H}_{\mathcal{N}(v)}(t)
\,\middle|\,
\mathcal{H}_v(t),\mathbf{z}_v
\right)
}_{\substack{
I_v^{(\mathrm{prop})}: \text{propagation term}
}} .
\end{aligned}
\label{eq:information_decomposition}
\end{equation}
Here, $\mathcal{H}_{\mathcal{N}(v)}(t)$ collects the historical feature sequences of node $v$'s neighbors. The terms $I_v^{(\mathrm{conf})}$, $I_v^{(\mathrm{local})}$, and $I_v^{(\mathrm{prop})}$ quantify fault-risk information from static configuration, local degradation history, and neighboring histories, respectively.

We convert this statistically averaged mutual information into sample-specific fault information gain using pointwise mutual information:
\begin{equation}
\Delta s_v^F(t)=\Delta s_v^{\mathrm{conf}}+\Delta s_v^{\mathrm{local}}(t)+\Delta s_v^{\mathrm{prop}}(t).
\end{equation}

Because the complete histories $\mathcal{H}_v(t)$ and $\mathcal{H}_{\mathcal{N}(v)}(t)$ are high-dimensional and costly to model, we compress them into the abnormal information carried by fault events: 
\begin{equation}
\Delta s_v^F(t;\mathcal{H})=\Delta s_v^F(t;\mathcal{A})+\zeta_v(t),
\end{equation}
where the compression residual is
\begin{equation}
\zeta_v(t)=
\log\frac{P\!\left(F_v(t+1)=1\mid \mathcal{H}_v(t),\mathcal{H}_{\mathcal{N}(v)}(t),\mathbf{z}_v\right)}
{P\!\left(F_v(t+1)=1\mid \mathcal{A}_v(t),\mathcal{A}_{\mathcal{N}(v)}(t),\mathbf{z}_v\right)}.
\end{equation}
Here, $\zeta_v(t)$ directly measures the information lost by replacing full feature histories with event histories. When $\zeta_v(t)\approx 0$, the fault-event sets preserve the relevant historical information:
\begin{equation}
\begin{aligned}
&P\!\left(F_v(t+1)=1\mid\mathcal{H}_v(t),\mathcal{H}_{\mathcal{N}(v)}(t),\mathbf{z}_v\right)\\
&\quad=P\!\left(F_v(t+1)=1\mid\mathcal{A}_v(t),\mathcal{A}_{\mathcal{N}(v)}(t),\mathbf{z}_v\right).
\end{aligned}
\end{equation}

To obtain a graph propagation form, we decompose the joint event-history contribution into source-node contributions:
\begin{equation}
\Delta s_v^{\mathrm{local}}(t)+\Delta s_v^{\mathrm{prop}}(t)
=\sum_{u\in\mathcal{N}(v)\cup\{v\}}\Delta s_{u\to v}(t)+\eta_v(t),
\end{equation}
where $\eta_v(t)$ captures higher-order synergy and redundancy among sources. When them are negligible, i.e., $\eta_v(t)\approx 0$, the gain becomes the sum of source-wise contributions:
\begin{equation}
\Delta s_v^{\mathrm{local}}(t)+\Delta s_v^{\mathrm{prop}}(t)
=\sum_{u\in\mathcal{N}(v)\cup\{v\}}\Delta s_{u\to v}(t).
\end{equation}

\textbf{Theorem 4 (Atomic integral property of a marked counting measure).} Let $\delta_{x_k}$ be the Dirac measure at event $x_k$ and $\psi(x)$ an event-response kernel. Then
\begin{equation}
\int \psi(x)\sum_k\delta_{x_k}(dx)=\sum_k\psi(x_k).
\end{equation}

Represent source node $u$'s fault history by the marked counting measure
\begin{equation}
N_u(d\xi,da)=
\sum_{\substack{t_k^u\leq t\\F_u(t_k^u)=1}}
\delta_{(t_k^u,S_u(t_k^u))}(d\xi,da),
\end{equation}
which retains event time and anomaly severity as marks. The cumulative gain transmitted from source $u$ to target $v$ is
\begin{equation}
\Delta s_{u\to v}(t)=
\int_{[0,t)\times\mathbb{R}_{+}}
\psi_{u\to v}(a,t-\xi)N_u(d\xi,da).
\end{equation}
By Theorem~4, its event-wise form is
\begin{equation}
\Delta s_{u\to v}(t)=
\sum_{\substack{t_k^u\leq t \\F_u(t_k^u)=1}}
\psi_{u\to v}\!\left(S_u(t_k^u),t-t_k^u\right).
\end{equation}
Consequently, the total fault information gain is
\begin{equation}
\begin{aligned}
\Delta s_v^F(t)=&\Delta s_v^{\mathrm{conf}}+
\\&\sum_{u\in\mathcal{N}(v)\cup\{v\}}\sum_{\substack{t_k^u\leq t\\F_u(t_k^u)=1}}
\psi_{u\to v}\!\left(S_u(t_k^u),t-t_k^u\right).
\end{aligned}
\end{equation}

For an interpretable, estimable model, we decompose each event's risk information as
\begin{equation}
\psi_{u\to v}\!\left(S_u(t_k^u),t-t_k^u\right)
=W_{u\to v}\,\varphi\!\left(S_u(t_k^u)\right)\,g_{u\to v}\!\left(t-t_k^u\right),
\end{equation}
where $W_{u\to v}$ is the learned propagation weight, $\varphi(S_u(t_k^u))=1-e^{-S_u(t_k^u)}$ is a bounded severity function, and $g_{u\to v}(\cdot)$ is a temporal-decay kernel. This factorization encodes principles: stronger propagation paths and more severe faults contribute more risk information, whereas older events exert less.

Let $\Phi_{u\to v}(t)$ denote the effective fault-information stock transmitted from $u$ to $v$, with dynamics
\begin{equation}
\begin{aligned}
\frac{d\Phi_{u\to v}(t)}{dt}
&=-\gamma_{u\to v}\Phi_{u\to v}(t)
+\\&\sum_{\substack{t_k^u\leq t\\F_u(t_k^u)=1}}
W_{u\to v}\varphi\!\left(S_u(t_k^u)\right)\delta(t-t_k^u),
\end{aligned}
\end{equation}
where $\gamma_{u\to v}>0$ is the learned memory-decay rate. The solution is
\begin{equation}
\Phi_{u\to v}(t)=
\sum_{\substack{t_k^u\leq t\\F_u(t_k^u)=1}}
W_{u\to v}\varphi\!\left(S_u(t_k^u)\right)
e^{-\gamma_{u\to v}(t-t_k^u)},
\end{equation}
corresponding to the exponential-decay kernel
\begin{equation}
g_{u\to v}(t-t_k^u)=e^{-\gamma_{u\to v}(t-t_k^u)}.
\end{equation}
Substitution yields the history-based fault information gain
\begin{equation}
\begin{aligned}
&\Delta s_v^F(t)
=\Delta s_v^{\mathrm{conf}}+\\&\sum_{u\in\mathcal{N}(v)\cup\{v\}}W_{u\to v}\sum_{\substack{t_k^u\leq t\\F_u(t_k^u)=1}}
\left(1-e^{-S_u(t_k^u)}\right)e^{-\gamma_{u\to v}(t-t_k^u)}.
\end{aligned}
\end{equation}
Because fault probability increases with information gain, we map the gain to a valid probability using the sigmoid function:
\begin{equation}
\begin{aligned}
& P\!\left(
F_v(t+1)=1
\,\middle|\,
\mathcal{H}_v(t),
\mathcal{H}_{\mathcal{N}(v)}(t),
\mathbf{z}_v
\right)                                                   \\
&\quad = \sigma\!\left(\Delta s_v^{F}(t)\right)            \\
&\quad = \sigma\Biggl(
    \mu_v\!\left(\mathbf{z}_v\right)
    + \sum_{u\in\mathcal{N}(v)\cup\{v\}}
      W_{u\rightarrow v}                              \\
&\qquad\qquad {}\times
      \sum_{\substack{
          t_k^u\leq t\\
          F_u(t_k^u)=1
      }}
      \left(1-e^{-S_u(t_k^u)}\right)
      e^{-\gamma_{u\to v}(t-t_k^u)}
\Biggr).
\end{aligned}
\label{eq:failure_probability}
\end{equation}
where $\mu_v(\mathbf{z}_v)$ maps static configuration $\mathbf{z}_v$ to background abnormal information.

\section{FEMNet}
\label{sec:femnet}
Building on IEFP, FEMNet couples conditional diffusion with a fault-aware denoising network to predict KPI distributions under complex fault scenarios. This section presents its overall architecture and details denoising mechanism.

\begin{figure*}
\centerline{\includegraphics[width=0.9\textwidth]{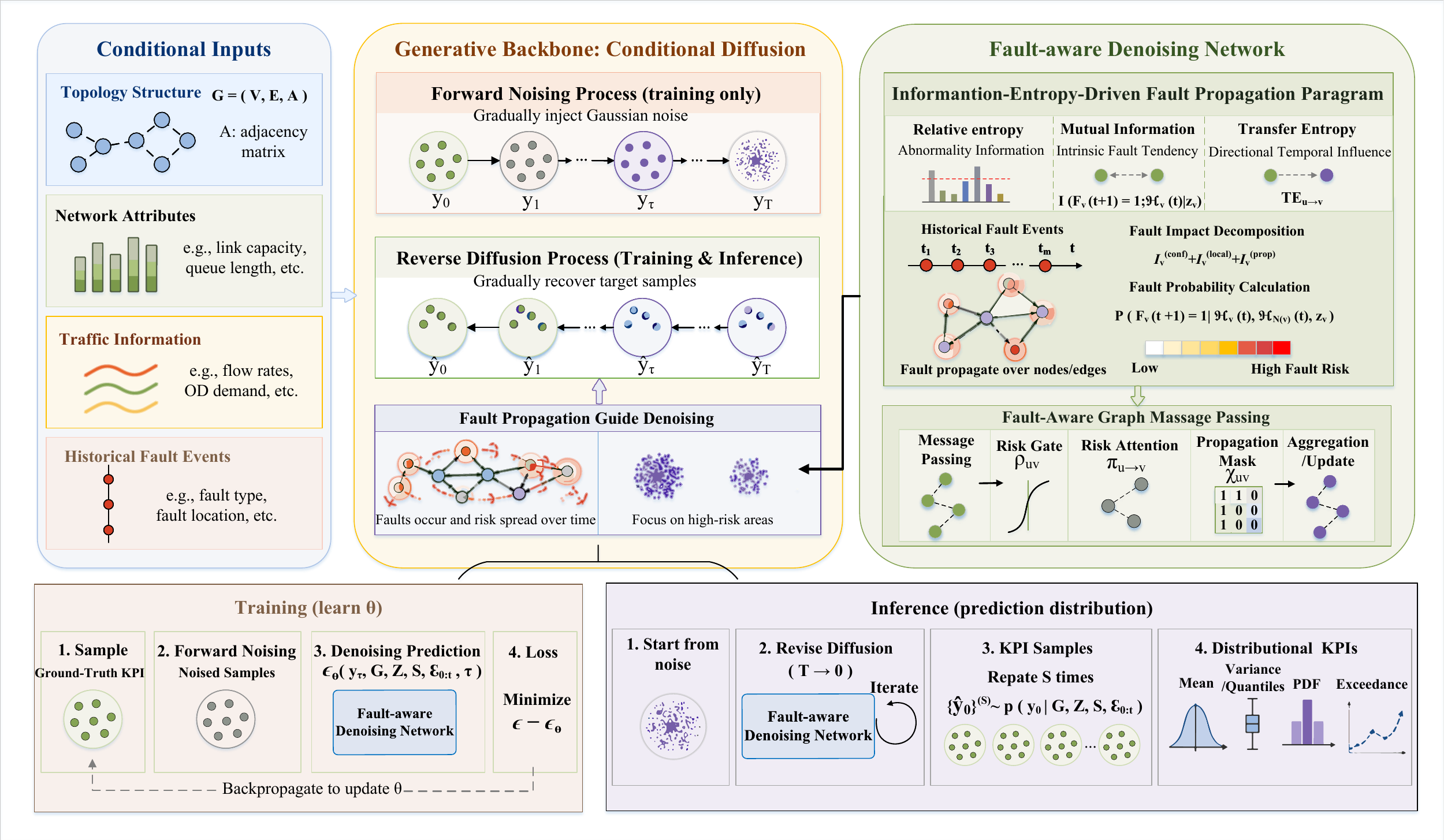}}
\caption{FEMNet architecture and end-to-end workflow. Upper blocks trace conditional information through the diffusion backbone and fault-aware denoising network, whereas lower blocks distinguish model training (left) from repeated-sampling inference for predictive KPI distributions (right).}
\label{fig:FEMNet_1}
\end{figure*}

\subsection{Model Overview}
Fig.~\ref{fig:FEMNet_1} shows FEMNet's three components: conditional inputs, a generative backbone, and a fault-aware denoising network. The inputs comprise topology structure, network attributes, traffic information, and historical fault event sequences. Conditional diffusion generates probabilistic KPI distributions rather than deterministic point estimates. Fault-aware graph message passing mechanism injects propagation context into denoising. 

The conditional diffusion backbone models the target KPI as a random variable and learns its probabilistic distribution. During training, the forward noising process gradually perturbs ground-truth KPI samples. FEMNet learns the reverse process conditioned on network and fault information. During inference, FEMNet starts from Gaussian noise and iteratively denoises it into conditional KPI samples. Repeated sampling approximates the future KPI distribution, capturing its fluctuation range, uncertainty level, and shape under fault scenarios.

A key design of FEMNet lies in its fault-aware denoising network. IEFP first maps historical fault event sequences to fault risks over nodes and links. As dynamic propagation context, these risks guide graph message passing through risk gates, attention weights, and propagation masks. Consequently, the reverse process captures both the current network snapshot and the historical propagation and cumulative effects of faults.

\subsection{Fault-Aware Denoising Network}
The fault-aware denoising network serves as the reverse-process predictor, which effectively model the fault impact. At step $\tau$, it predicts the noise added to KPI sample $\mathbf{y}_\tau$:
\begin{equation}
\hat{\boldsymbol{\epsilon}}_\tau
=\boldsymbol{\epsilon}_\theta(\mathbf{y}_\tau,\tau,\mathcal{C}).
\label{eq:femnet_noise_prediction}
\end{equation}
The corresponding clean-KPI estimate is
\begin{equation}
\hat{\mathbf{y}}_0
=\frac{\mathbf{y}_\tau
-\sqrt{1-\bar{\alpha}_\tau}\,
\hat{\boldsymbol{\epsilon}}_\tau}
{\sqrt{\bar{\alpha}_\tau}}.
\label{eq:clean_kpi_estimate}
\end{equation}
The predictor is evaluated at each reverse step, and the final sample is transformed back to the original KPI scale.

Using~\eqref{eq:failure_probability}, FEMNet calculates next-timestep node failure probabilities as fault risk intensity embedding $r_v$. we initialize
\begin{equation}
h_v^0=[\mathbf{x}_v(t),r_v,\varepsilon_k],
\label{eq:hidden_state_initialization}
\end{equation}
where $[\cdot]$ denotes concatenation. When $v$ is fault-free, $\varepsilon_k=\mathbf{0}$. $h_v^i$ is the learned fault-aware state after $i$ message-passing layers. We use $h_f^i$, $h_q^i$, and $h_l^i$ for flow, queue, and link nodes, respectively.

We first introduce risk gate $\rho_{uv}$ to control the effective amount of propagated message:
\begin{equation}
\rho_{uv}=\mathrm{MLP}([r_u,r_v]).
\end{equation}
Here, $\rho_{uv}$ combines the endpoint risk embeddings.

To capture heterogeneous influence across neighbors, we introduce risk attention. Attention bias $\pi_{u\to v}$ from $u$ to $v$ is 
\begin{equation}
\pi_{u\to v}=\frac{\exp{(\kappa_{u\to v})}}{\sum_{u^{\prime}\in\mathcal{N}(v)}\exp{(\kappa_{u^{\prime}\to v})}},
\end{equation}
where
\begin{equation}
\kappa_{u\to v}=w_a(h_u,h_v)+w_b(r_{u}).
\end{equation}
Here, $w_a$ and $w_b$ are learnable scoring functions. The attention mechanism emphasizes high-risk entities or edges when propagation is likely to affect the target. 

We further apply a dynamic propagation mask to suppress irrelevant messages under weak fault coupling. When cumulative risk or propagation relevance falls below a learnable threshold $\vartheta$, the corresponding edge is disabled. Let $\chi_{uv}\in\{0,1\}$ denote this mask:
\begin{equation}
\chi_{uv}=\begin{cases}1, & \rho_{uv}\geq\vartheta \\0, & \rho_{uv}<\vartheta\end{cases}.
\end{equation}
Thus, $\rho_{uv}$ controls message magnitude, while $\chi_{uv}$ removes weakly coupled propagation paths.

Propagation differs by edge type. $U_v(\cdot)$ is implemented as GRU. For flow-node updates, queue and link messages, modulated by risk gate and propagation  mask, are sequentially aggregated along the routing path. Namely,
\begin{equation}
m_{q\to f}^i=\chi_{fq}\rho_{fq}h_q^i,
\end{equation}
\begin{equation}
m_{l\to f}^i=\chi_{fl}\rho_{fl}h_l^i,
\end{equation}
\begin{equation}
h_{f}^{i+1}\leftarrow U_{f}(h_{f}^i,[m_{q\to f}^{i},m_{l\to f}^{i}]).
\end{equation}

Queue’s state depends on its traversing flows and associated link. Risk attention is incorporated into multiflow aggregation to capture flow-specific contributions. $U_q$ integrates queue features with risk-modulated flow and link aggregates to update the queue state:
\begin{equation}
m_q^i=\sum_{f\in\mathcal{N}(q)}\pi_{f\to q}\rho_{qf}h_f^i+\chi_{ql}\rho_{ql}h_l^i,
\end{equation}
\begin{equation}
h_{q}^{i+1}\leftarrow U_{q}(h_{q}^i,m_{q}^i).
\end{equation}

Link's state similarly depends on its traversing flows and connected queues, giving the message and update:
\begin{equation}
m_l^i=\sum_{f\in\mathcal{N}(l)}\pi_{f\to l}\rho_{lf}h_f^i+\chi_{lq}\rho_{lq}h_q^i,
\end{equation}
\begin{equation}
h_{l}^{i+1}\leftarrow U_{l}(h_{l}^i,m_{l}^i).
\end{equation}

Collectively, these edge-type-specific updates yield fault-conditioned representations for $\epsilon_\theta$, thereby linking the IEFP-derived risk context to conditional KPI distribution generation.

\section{Experiments}
This section first describes the dataset generation and experimental settings, then evaluates IEFP and FEMNet in fault prediction, deterministic and probabilistic KPI prediction, component ablations, and efficiency.

\subsection{DataSet Generation}
Our ns-3 simulations use the Abilene, GEANT and Germany50 topologies and corresponding traffic matrices from SNDlib\cite{orlowski2010sndlib}, a widely used data library for telecommunication network design and performance evaluation. We inject diverse faults and record propagation trajectories and KPI samples. After network failures, ns-3 simulates the redistribution of network resources and the resulting utilization changes. This process may establish a new equilibrium or trigger subsequent degradation and failure. Thus, the dataset captures both direct multi-fault impacts and cascading propagation effects.

The injected faults cover five categories: node/device faults (node or interface outages and degradation), link faults (disconnection, bandwidth reduction, delay, packet loss, and bit errors), control-plane faults (adjacency loss, slow convergence, route-cost changes, and flapping-induced loops or black holes), data-plane faults (congestion, queue overflow, and RED/WRED drops), and application-level faults (flash crowds and DDoS traffic). Combining these faults with different topologies, traffic conditions, and network configurations yields scenarios with diverse propagation patterns and performance impacts for evaluating IEFP and FEMNet.

\begin{table*}[htbp]
\centering
\caption{Deterministic KPI-prediction performance of FEMNet and the baselines.}
\label{tab:det_results}
\renewcommand{\arraystretch}{1.1}
\begin{tabular}{ccccccccccccc} 
\hline
\multicolumn{1}{l}{} & \multicolumn{4}{c}{Delay}                                                                 & \multicolumn{4}{c}{Loss-Ratio}                                                            & \multicolumn{4}{c}{Throughput}                                                             \\ 
\hline
\multicolumn{1}{l}{} & MSE $\downarrow$ & MAE $\downarrow$ & MAPE $\downarrow$ & R\textsuperscript{2} $\uparrow$ & MSE $\downarrow$ & MAE $\downarrow$ & MAPE $\downarrow$ & R\textsuperscript{2} $\uparrow$ & MSE $\downarrow$ & MAE $\downarrow$ & MAPE $\downarrow$ & R\textsuperscript{2} $\uparrow$  \\ 
\cline{2-13}
STGNN                & 0.4928           & 0.3278           & 66.53\%           & 0.7137                          & 0.6964           & 0.3848           & 113.01\%          & 0.7232                          & 0.6519           & 0.3844           & 99.47\%           & 0.8110                           \\
DGAT                 & 0.4139           & 0.3154           & 63.91\%           & 0.7595                          & 0.6163           & 0.3869           & 123.04\%          & 0.7551                          & 0.5714           & 0.3663           & 91.55\%           & 0.8344                           \\
xNet                 & 0.4281           & 0.3263           & 71.83\%           & 0.7512                          & 0.5870           & 0.3833           & 124.48\%          & 0.7667                          & 0.5715           & 0.3319           & 91.04\%           & 0.8343                           \\
EAGLE                & 0.3962           & 0.3069           & 63.85\%           & 0.7698                          & 0.4854           & 0.3448           & 113.42\%          & 0.8071                          & 0.5686           & 0.3999           & 80.93\%   & 0.8307                           \\
DCRNN                & 0.4028           & 0.3003           & 58.48\%   & 0.7659                          & 0.5263           & 0.3524           & 113.37\%          & 0.7908                          & 0.5637           & 0.2989   & 83.27\%           & 0.8366                           \\
Routenet-Fermi       & 0.3768   & 0.2809   & 61.30\%           & 0.7811                  & 0.4817   & 0.3368   & 101.81\%  & 0.8086                  & 0.5613   & 0.3002           & 83.74\%           & 0.8372                   \\ 
\hline
FEMNet               & \textbf{0.2282}  & \textbf{0.1678}  & \textbf{38.29\%}  & \textbf{0.8674}                 & \textbf{0.2814}  & \textbf{0.2147}  & \textbf{55.17\%}  & \textbf{0.8881}                 & \textbf{0.3627}  & \textbf{0.1868}  & \textbf{54.02\%}  & \textbf{0.8949}                  \\
\hline
\end{tabular}
\end{table*}

\begin{figure*}
	\centering
    \centerline{\includegraphics[width=0.8\textwidth]{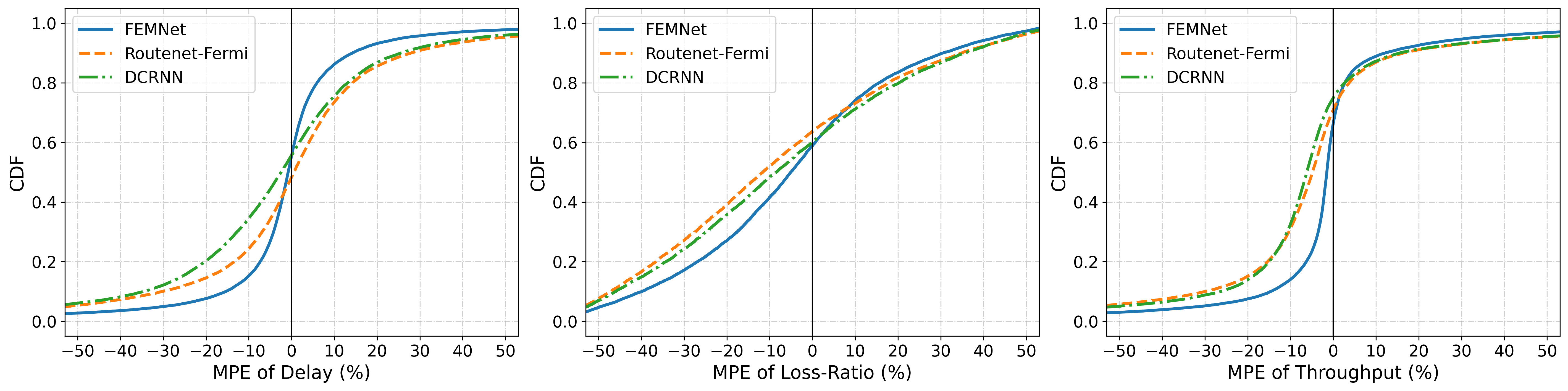}}
	\caption{Cumulative distribution functions (CDFs) of signed MPE for test-set predictions by FEMNet, RouteNet-Fermi, and DCRNN.}
	\label{fig:mpe}
\end{figure*}

\begin{table*}
\centering
\caption{Probabilistic delay-prediction performance of FEMNet and the baselines.}
\label{tab:distribution_results}
\renewcommand{\arraystretch}{1.05}
\begin{tabular}{ccccccccc} 
\hline
\multicolumn{1}{l}{} & \multirow{2}{*}{CRPS $\downarrow$} & \multirow{2}{*}{ES-norm $\downarrow$} & \multicolumn{3}{c}{90\% Prediction Interval}                  & \multicolumn{3}{c}{95\% Prediction Interval}                   \\ 
\cline{4-9}
\multicolumn{1}{l}{} &                                    &                                       & PICP $\rightarrow 90\%$ & MPIW $\downarrow$ & IS $\downarrow$ & PICP $\rightarrow 95\%$ & MPIW $\downarrow$ & IS $\downarrow$  \\ 
\hline
GNN-Denoise          & 0.4077                             & 0.6607                                & 0.8408                  & 2.2957            & 3.8813          & 0.9035                  & 2.9657            & 4.7136           \\
Rt-Denoise           & 0.2717                             & 0.5283                                & 0.8949                  & 1.7081            & 2.6046          & 0.9324                  & 2.1112            & 3.1674           \\
ReDiSC               & 0.2302                             & 0.4496                                & 0.8864                  & 1.0802            & 2.1667          & 0.9245                  & 1.3455            & 2.8388           \\
SimDiff              & 0.2088                             & 0.4044                                & \textbf{0.8989}         & 1.0887            & 1.8511          & 0.9354                  & 1.3297            & 2.3755           \\
LGD                  & 0.1577                     & 0.3336                        & 0.9031          & 0.8576    & 1.4708  & 0.9367          & 1.0683    & 1.8917   \\ 
\hline
FEMNet               & \textbf{0.1305}                    & \textbf{0.2919}                       & 0.9068                  & \textbf{0.6451}   & \textbf{1.2640} & \textbf{0.9432}         & \textbf{0.7903}   & \textbf{1.7043}  \\
\hline
\end{tabular}
\end{table*}

\begin{table}
\centering
\caption{Fault prediction performance of IEFP and the baselines.}
\label{tab:results}
\renewcommand{\arraystretch}{1.05}
\begin{tabular}{cccccc} 
\hline
\multicolumn{1}{l}{} & Accuracy        & TNR                        & Precision       & Recall          & F1 score         \\ 
\hline
LSTM                 & 0.7149          & \multicolumn{1}{l}{0.7961} & 0.5653          & 0.6176          & 0.5542           \\
GNN                  & 0.7915          & \multicolumn{1}{l}{0.8311} & 0.6822          & 0.6621          & 0.6628           \\
Causal model         & 0.8030          & 0.8369                     & 0.6980          & 0.6949          & 0.6506           \\
HIC model            & 0.8324          & 0.8544                     & 0.7099          & \textbf{0.8557} & 0.7319           \\
GPIAN                & 0.8813  & 0.9466             & 0.8673  & 0.7550          & 0.8085   \\ 
\hline
IEFP                 & \textbf{0.9225} & \textbf{0.9704}            & \textbf{0.9162} & 0.8159  & \textbf{0.8494}  \\
\hline
\end{tabular}
\end{table}

\begin{figure}[!t]
	\centering
	\includegraphics[width=\columnwidth]{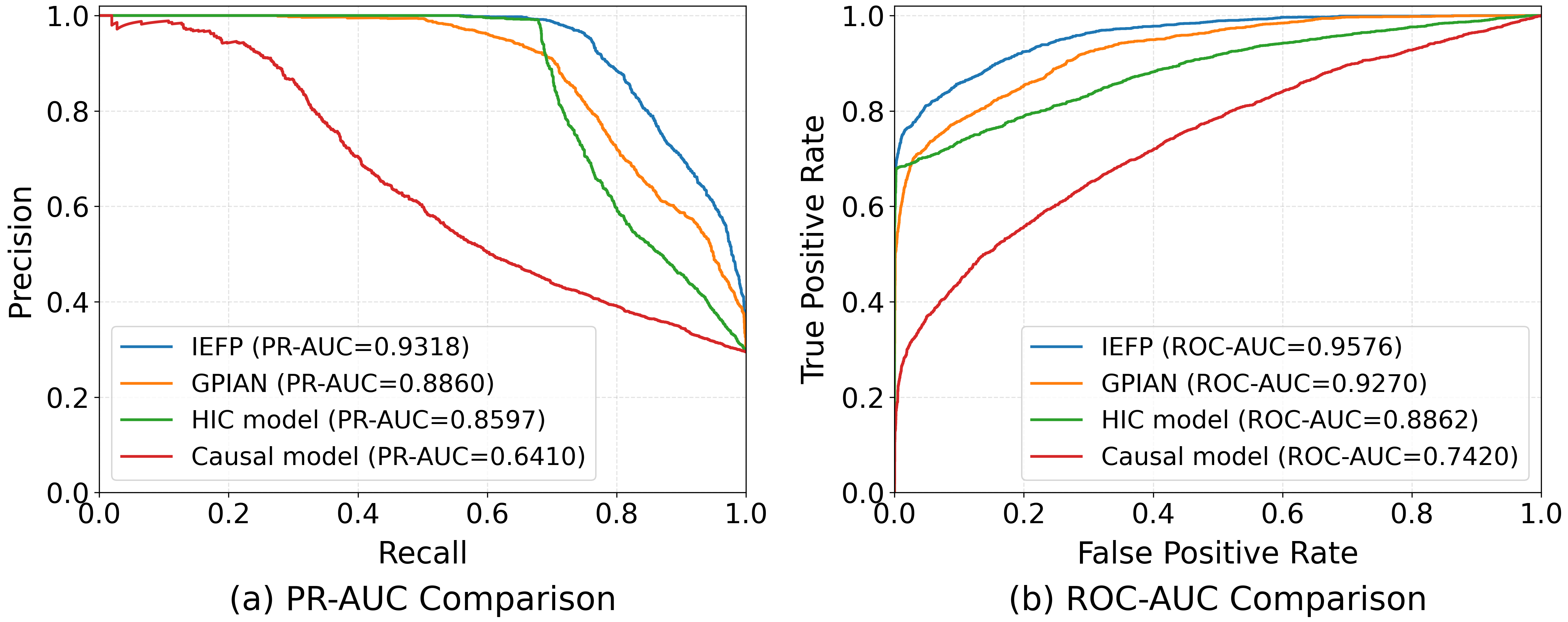}
	\caption{Precision--recall (PR) curves and Receiver operating characteristic (ROC) curves for fault prediction by IEFP and baselines. Legend entries report the corresponding areas under the curves (PR-AUCs and ROC-AUCs).}
	\label{fig:auc}
\end{figure}

\subsection{Experimental Settings}
We split the dataset into training, validation, and test sets at 60\%:20\%:20\%. All methods are trained with Adam for 500 epochs, a learning rate of 0.001 and a batch size of 16, and the checkpoint with the best validation performance is selected. All diffusion models employ 100 denoising steps, and we generate 50 KPI samples per test instance. For a fair comparison, all methods use identical input features, including fault information. Experiments are implemented in PyTorch on Intel Xeon Silver 4410Y CPU and NVIDIA RTX A6000 GPU.

We compare FEMNet with representative learning-based methods to access deterministic performance prediction and evaluate IEFP against GPIAN, HIC model, Causal model, and related fault-prediction methods. For probabilistic prediction, we compare FEMNet with the denoising-backbone baselines, GNN-Denoise and Rt-Denoise (RouteNet-Erlang), and diffusion models, including ReDiSC, SimDiff, and LGD. We further ablate the main components of FEMNet and IEFP, and profile memory and runtime costs. 

Deterministic metrics are mean squared error (MSE), mean absolute error (MAE), mean absolute percentage error (MAPE), mean percentage error (MPE) and Coefficient of Determination ($R^2$). Fault prediction is assessed via accuracy, precision, recall, F1 score, and true negative rate (TNR). Probabilistic metrics are continuous ranked probability score (CRPS), normalized energy score (ES-norm), prediction interval coverage probability (PICP), mean prediction interval width (MPIW), and interval score (IS) for 90\% and 95\% prediction intervals (PIs). We also report inference and training time, GPU memory, and parameter count. 

\begin{figure*}[!t]
	\centering
\includegraphics[width=0.72\textwidth]{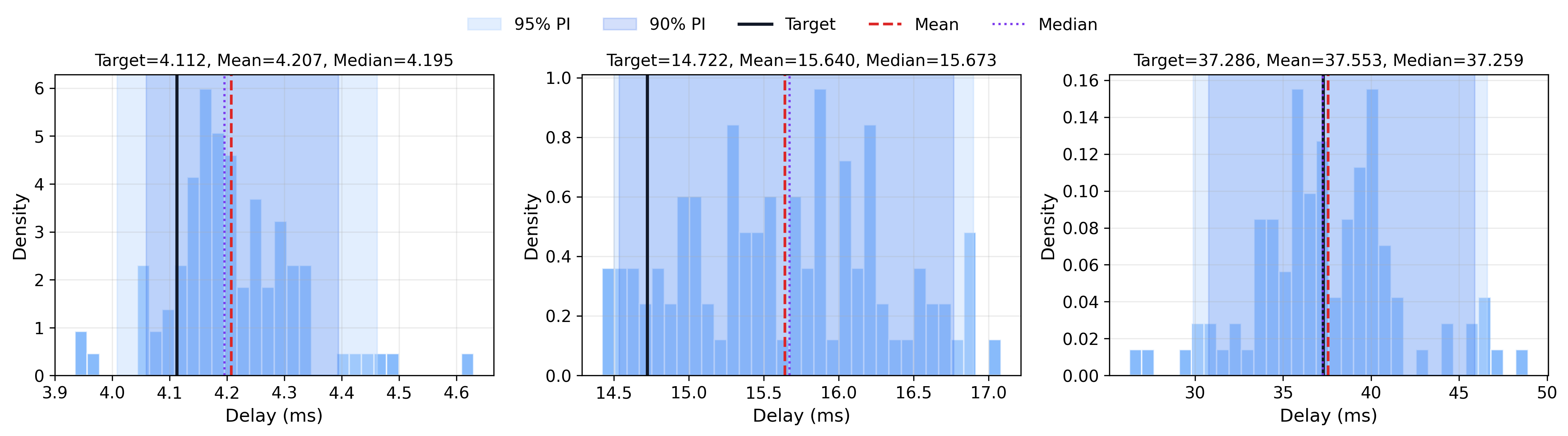}
	\caption{Predictive delay distributions produced by FEMNet for three representative instances with increasing target values from left to right.}
	\label{fig:distribution_samples}
\end{figure*}

\begin{table*}
\centering
\caption{Efficiency comparison of FEMNet and the representative baselines for delay prediction.}
\label{tab:inference}
\renewcommand{\arraystretch}{1.05}
\begin{tabular}{ccccccc} 
\hline
\multicolumn{1}{l}{}      & \multirow{2}{*}{IEFP} & \multicolumn{3}{c}{FEMNet}                                              & \multirow{2}{*}{\begin{tabular}[c]{@{}c@{}}LGD\\(T=100,S=8)\end{tabular}} & \multirow{2}{*}{Routenet-Fermi}  \\ 
\cline{3-5}
\multicolumn{1}{l}{}      &                       & T=1,S=1 & \multicolumn{1}{l}{T=100,S=1} & \multicolumn{1}{l}{T=100,S=8} &                                                                           &                                  \\ 
\hline
GPU Memory Usage (MB)     & 17.42                 & 30.36   & 30.36                         & 30.36                         & 38.72                                                                     & 29.22                            \\
Inference Time (ms/batch) & 0.848                 & 10.13   & 848                           & 5947                          & 2956                                                                      & 9.76                             \\
Training Time (ms/epoch)  & 3172                  & 6657    & 6871                          & 7003                          & 4691                                                                      & 5564                             \\
Parameter Count           & 13764                 & 199370  & 199370                        & 199370                        & 295175                                                                    & 105985                           \\
\hline
\end{tabular}
\end{table*}

\begin{table}
\centering
\caption{FEMNet ablations. The w/o Fault, Mod and Diff variants remove fault information, fault-aware message mechanism, and conditional diffusion, respectively}
\label{tab:ablation}
\renewcommand{\arraystretch}{1.05}
\begin{tabular}{ccccccc} 
\hline
\multicolumn{1}{l}{} & \multicolumn{2}{c}{Delay} & \multicolumn{2}{c}{Loss-Ratio} & \multicolumn{2}{c}{Throughput}  \\ 
\hline
\multicolumn{1}{l}{} & MSE    & MAE              & MSE    & MAE                   & MSE    & MAE                    \\ 
\cline{2-7}
w/o Fault            & 0.5352 & 0.3813           & 0.7128 & 0.4264                & 0.6738 & 0.3568                 \\
w/o Mod              & 0.3481 & 0.2381           & 0.4460 & 0.2763                & 0.5376 & 0.3250                 \\
w/o Diff             & 0.2551 & 0.1855           & 0.3125 & 0.2258                & 0.3941 & 0.1978                 \\ 
\hline
FEMNet               & 0.2282 & 0.1678           & 0.2814 & 0.2147                & 0.3627 & 0.1868                 \\
\hline
\end{tabular}
\end{table}

\begin{table}
\centering
\caption{IEFP ablations. The w/o $W_{\mathrm{prop}}$, Severity and Config variants remove edge-specific propagation weight, entropy-derived severity, and configuration term, respectively. Shared Decay replaces edge-specific decay rates with single learnable rate.}
\label{tab:ablation_IEFP}
\renewcommand{\arraystretch}{1.05}
\begin{tabular}{cccccc} 
\hline
\multicolumn{1}{l}{} & Accuracy & TNR    & Precision & Recall & F1 score  \\ 
\hline
w/o $W_\text{prop}$  & 0.8766   & 0.9276 & 0.8212    & 0.7871 & 0.7973    \\
w/o~Severity         & 0.8896   & 0.9419 & 0.8729    & 0.7618 & 0.8069    \\
w/o Config           & 0.9013   & 0.9422 & 0.8865    & 0.7844 & 0.8155    \\
Shared Decay         & 0.9027   & 0.9468 & 0.8947    & 0.7915 & 0.8246    \\ 
\hline
IEFP                 & 0.9225   & 0.9704 & 0.9162    & 0.8159 & 0.8494    \\
\hline
\end{tabular}
\end{table}

\subsection{Evaluation Results}
\textbf{Deterministic performance Prediction.}  
Table~\ref{tab:det_results} compares FEMNet with deterministic prediction baselines on three KPIs. FEMNet consistently achieves the best performance for every KPI–metric pair. Compared with the best-performing baseline for each pair, FEMNet achieves average reductions of 38.6\%, 37.9\%, and 38.9\% in MSE, MAE, and MAPE, respectively, and an average improvement of 9.2\% in $R^2$ across the three KPIs. Fig.~\ref{fig:mpe} further shows that FEMNet generally produces more concentrated signed-MPE errors than RouteNet-Fermi and DCRNN. FEMNet's closer concentration around zero indicates less systematic over- or underprediction. These results show that FEMNet substantially improves deterministic KPI prediction with propagation modeling, fault-aware graph reasoning, and conditional diffusion.

\textbf{Fault Prediction.} 
Table~\ref{tab:results} shows that IEFP achieves the highest accuracy (0.9225), TNR (0.9704), precision (0.9162), and F1 score (0.8494), with the second-highest recall (0.8159). Relative to the strongest overall baseline GPIAN, IEFP increases all indicators. The HIC model has the highest recall (0.8557), but its lower precision and F1 score indicate more false alarms. The causal model exhibits a less favorable precision-recall balance. Figs.~\ref{fig:auc} provide threshold-independent support: IEFP obtains the highest PR-AUC (0.9318) and ROC-AUC (0.9576). These results show that IEFP more reliably anticipates future faults while maintaining a favorable balance between detection and false alarms.

\textbf{Probabilistic Performance Prediction.} 
Table~\ref{tab:distribution_results} reports that FEMNet achieves the lowest CRPS (0.1305) and ES-norm (0.2919), reducing by 17.3\% and 12.5\% against LGD, respectively. For the 90\% PI, SimDiff has the PICP closest to nominal; FEMNet remains near nominal at 0.9068 and has lowest MPIW (0.6451) and IS (1.2640). For the 95\% PI, FEMNet has the closest PICP (0.9432) and lowest MPIW (0.7903) and IS (1.7043). Although FEMNet is not closest to nominal at 90\%, its narrower interval and lower score demonstrate better sharpness. Overall, FEMNet improves distributional accuracy without relying on excessively wide, conservative intervals for coverage. Fig.~\ref{fig:distribution_samples} visualizes three representative delay cases. Each target lies within both the 90\% and 95\% PIs, while the predictive mean and median remain consistent with the target scale. The low-delay case is relatively concentrated, whereas the higher-delay cases are broader and more dispersed, which illustrate instance-dependent, heteroscedastic uncertainty.

\textbf{Ablation Studies.} 
As shown in Table~\ref{tab:ablation}, removing fault information causes the largest errors, which demonstrates fault context is essential for understanding KPI degradation. Without fault-aware message passing, MSE and MAE increase by 52.7\% and 47.4\%, respectively, which verifies that fault features alone cannot replace risk-modulated graph reasoning. Conditional diffusion provides smaller but consistent gains for modeling uncertainty and complex latent distributions. Table~\ref{tab:ablation_IEFP} shows that edge-specific propagation weights are IEFP's most influential component: removing them lowers accuracy, TNR, precision, and F1 score by 4.59, 4.28, 9.50, and 5.21 percentage points, respectively, indicating overly broad propagation. Entropy-derived severity removal most harms recall because minor and severe fault events become less distinguishable. Configuration term removal degrades all metrics, which supports background-vulnerability modeling. Shared decay has the smallest impact, suggesting temporal heterogeneity provides a modest marginal contribution.

\textbf{Model Efficiency.} 
Let $N$ and $M$ denote the numbers of nodes and edges, respectively. Let $d$, $L$, $K$, $T$, $S$, and $P$ denote the hidden feature dimension, message-passing depth, historical fault count, reverse diffusion step count, inference sample count, and target-KPI dimension, respectively. The time complexity is $\mathcal{O}\!\left(KM+ST[L(Nd^2+Md)+P]\right)$ and space complexity is $\mathcal{O}(K+Nd+M+P)$. Table~\ref{tab:inference} shows that IEFP is lightweight, using 17.42 MB and 0.848 ms/batch. At $T=S=1$, FEMNet uses 30.36 MB and 10.13 ms/batch, close to RouteNet-Fermi, indicating limited cost from fault propagation encoding and graph reasoning. Increasing $T$ or $S$ raises inference latency nearly linearly, consistent with multi-step diffusion and repeated sampling. Memory and parameter count remain fixed because the denoiser is reused. Training time increases slightly because each instance samples one diffusion step and evaluates the denoiser once. At $(T,S)=(100,8)$, FEMNet trains and infers more slowly than LGD but uses less memory and fewer parameters. Thus, FEMNet trades between overhead and performance.

\section{Conclusion}
This paper studied probabilistic network performance prediction under complex faults. IEFP models fault propagation as abnormal information transmission over networks, and fault-aware graph message passing injects predictive fault risk into representation learning. FEMNet converts deterministic KPI prediction into conditional distribution generation. Experiments showed improved fault prediction, deterministic and probabilistic KPI prediction; ablations confirmed each major component, while efficiency tests demonstrated lightweight IEFP and controllable FEMNet overhead. This study assumes closed-set fault types and relies on supervised historical fault events and KPI observations, whereas practical labels may be sparse, noisy, or delayed. Future work will explore open-set faults, cross-domain transfer, label-efficient learning, and online adaptation.

\section*{Acknowledgments}
This work is supported by Hunan Provincial
Natural Science Foundation (Grant no.2025JJ90177) and Jiangxi Provincial Natural Science Foundation (Grant no.20253BAC280098).

\bibliographystyle{IEEEtran}
\bibliography{ref}

\end{document}